\documentstyle[prl,aps]{revtex}
\begin{document}
\draft
%next two lines for two-column version
\twocolumn\hsize\textwidth\columnwidth\hsize\csname
@twocolumnfalse\endcsname
\title{Spin gap in a quasi--one--dimensional  $\rm \bf S =\case{1}{2}$ antiferromagnet: Cu$_2$(1,4-Diazacycloheptane)$_2$Cl$_4$}
\author{Philip R.\ Hammar, Daniel H.\ Reich, and Collin Broholm}
\address{ Department of Physics and Astronomy, The Johns Hopkins 
University, Baltimore, MD 21218
}
\author{Frans Trouw}
\address{IPNS, Argonne National Laboratories, Argonne, IL 60439
}
\date{\today}
\maketitle
\begin{abstract}
Cu$_{2}$(1,4-diazacycloheptane)$_{2}$Cl$_{4}$ contains double chains of
spin--$\frac{1}{2}$ Cu$^{2+}$ ions.  We report ac susceptibility,
specific heat, and inelastic neutron scattering measurements on this
material. The magnetic susceptibility, $\chi (T)$, shows a rounded
maximum at $T=8$ K indicative of a low dimensional antiferromagnet with
no zero field magnetic phase transition. We compare the  $\chi (T)$
data to exact diagonalization results for various one dimensional spin
Hamiltonians and find excellent agreement for a spin ladder with
intra--rung coupling $J_1 = 1.143(3)$ meV and two mutually frustrating
inter--rung interactions:  $J_2 = 0.21(3)$ meV and $J_3 = 0.09(5)$
meV.  The specific heat in zero field is exponentially activated with an
activation energy $\Delta = 0.89(1)$ meV.  A spin gap is also found
through inelastic neutron scattering  on powder samples which identify
a band of magnetic excitations for $0.8<\hbar\omega<1.5$ meV.  Using
sum--rules we derive an expression for the dynamic spin correlation
function associated with non-interacting propagating triplets in a spin
ladder. The van-Hove singularities of such a model are not observed in
our scattering data indicating that magnetic excitations in
Cu$_{2}$(1,4-diazacycloheptane)$_{2}$Cl$_{4}$ are more complicated. For
magnetic fields above $H_{c1} \simeq 7.2$ T specific heat data versus
temperature show anomalies indicating a phase transition to an ordered
state below $T = 1 $ K.
\end{abstract}
\pacs{75.50.Ee, 75.40.-s}
%next two lines for two-column version
%\narrowtext
%\vskip2pc]

\section{Introduction}
As a consequences of its quantum criticality, the low-temperature
properties of the Heisenberg spin--1/2 
antiferromagnetic chain   change dramatically
with small perturbations.  
For example,  weak coupling between a macroscopic set of
spin chains leads to N\'{e}el order at $T=0$ while modifications to the
one dimensional network of spin interactions can induce spin--gaps and
suppress N\'{e}el order.  The gapped phases of one-dimensional
spin--1/2 systems are simple examples of a qualitatively new type of
magnetism where the many-body eigenstates are 
more conveniently described in terms  of the singlets and triplets
of near-neighbor coupled spin pairs,
rather than in terms of the Zeeman states of individual
spins.

Theorists have established that spin gaps can be induced by
bond alternation,\cite{Bonner79,Bonner83} 
by frustration due to next-nearest neighbor 
interactions, \cite{Shastry81,MajumdarGhosh69,white}
and by coupling even numbers of chains to form spin 
ladders.\cite{dagotto,LadderTheory} Spin--gaps have been observed in the
quasi--one--dimensional spin--1/2 systems $\rm Cu(NO_3)_2\cdot 2.5
H_2O$ \cite{Bonner83,Eckert}, $\rm CuGeO_3$ \cite{arai,ain},
SrCu$_2$O$_3$  \cite{Azuma94}, $\rm (La,Sr,Ca)_{14}Cu_{24}O_{41}$,
\cite{Carter96}
Cu$_2$(1,4-Diazacycloheptane)$_2$Cl$_4$ (CuHpCl)
\cite{Chiari90,Hammar,Chaboussant,cuhpclnmr}, and $\rm (VO)_2P_2O_7$
\cite{Johnston87,eccleston,garrettprb,garrettprl}.  Apart from the 
spin--Peierls system $\rm CuGeO_3$,
all of these other materials were originally thought to be spin ladders. 
However, subsequent experiments established that $\rm Cu(NO_3)_2\cdot 2.5
H_2O$ and $\rm (VO)_2P_2O_7$ actually consist of alternating spin--chains which
are perpendicular to the putative
spin-ladder. \cite{garrettprb,garrettprl} This highlights the difficulty
of distinguishing between different Hamiltonians with spin gaps on the
basis of measurements which mainly probe the magnetic density of
states.

This paper is an experimental study of the gapped spin--1/2 system 
CuHpCl. Because  magnetic exchange
interactions in this material are quite small it is an
excellent system in which to study the
field and temperature dependence
of spin correlations in a gapped spin--1/2 system.\cite{chitra97}
The Cu dimer formula units in CuHpCl stack to form double
chains. Correspondingly  the magnetic interactions
between Cu spins are thought to yield a spin ladder as shown in 
Fig.~\ref{structure} with a spin--Hamiltonian of the form
\begin{equation} 
{\cal H}  =  J_1 \sum_i {\bf S}_{2i} \cdot {\bf S}_{2i+1} 
 + J_2 \sum_i ({\bf S}_{2i} \cdot {\bf S}_{2i+2} + 
           {\bf S}_{2i-1} \cdot {\bf S}_{2i+1})
 + J_3 \sum_i{\bf S}_{2i-1} \cdot {\bf S}_{2i}  . 
\label{H_NNN} 
\end{equation} 
We report susceptibility, specific heat and neutron scattering
experiments which determine the spin gap and magnetic density of states
in CuHpCl.  While our susceptibility and specific heat data can be
interpreted in terms of a spin ladder, details of the inelastic magnetic
neutron scattering data are inconsistent with a simple approximation to the 
dynamic spin correlation function for a  spin ladder.

\section{Experimental Techniques}

CuHpCl is monoclinic, with space group $P 2_1/c$ and lattice constants $a 
= 13.406(3)$ \AA, $b = 11.454(2)$ \AA, $c = 12.605(3)$ \AA\ and $\beta 
= 115.01(2)^{\circ}$. \cite{Chiari90} The CuHpCl molecules stack to form 
chains in the (101) direction.  
The intradimer Cu--Cu distance is $d = $ 3.422 \AA\, and the
interdimer spacing along the chains is $u =$ 7.00\AA. We have grown
single crystals with  typical masses of 5-10 mg by slow cooling of 
saturated methanol 
solutions.  Crystals of this size were used for
ac susceptibility and specific heat measurements. The zero-field ac 
susceptibility for $T > 2 $ K was measured in a commercial 
SQUID magnetometer.  The heat capacity was measured in a dilution 
refrigerator using the relaxation method \cite{Bachmann72} in
magnetic fields up to 9 Tesla.

Inelastic neutron scattering measurements on a powder sample were
performed using the QENS time-of-flight spectrometer at the Intense
Pulsed Neutron Source (IPNS)  of Argonne National Laboratory.  
The sample consisted of 5.75 g of  powder in an annular aluminum
sample can with inner diameter 0.3 cm, outer diameter 1.0 cm, and height 
$12.75$ cm.  These dimensions were chosen to match the QENS  beam size,
while keeping the transmission through the hydrogenous sample  above
50\%.  

QENS is an inverse geometry time of flight spectrometer operating at a
fixed final energy $E_f$ = 3.635 meV selected by Bragg reflection from
three horizontally-focusing  pyrolytic graphite PG(002) analyzer
systems. The resulting elastic energy resolution is
$\delta E = 0.12$ meV Full Width at Half Maximum (FWHM).  Each channel is also equipped with a cooled BeO filter which
rejects neutrons with $E_f > 3.7$ meV through Bragg diffraction. The
analyzers can rotate about the sample on a common rotation stage whose
angular setting, $v$ can vary from $0$ to 30$^o$. The mean
scattering angles, $2\theta$, of the three analyzer systems are $v+25^o$,
$105^o - v$, and $v + 115^o$ respectively, and the angular acceptances
$\delta 2\theta$ are $7.5^o, 15^o$, and $15^o$. 

The magnetic neutron scattering cross section of a powder sample 
at wave vector transfer, $Q$, and energy
transfer, $\hbar\omega$ can be written\cite{lovesey}
\begin{equation}
\frac{{\rm d}^2\sigma}{{\rm d}\Omega{\rm d}E^{\prime}}=
\frac{k^{\prime}}{k}N r_0^2 |\frac{g}{2}f({\bf Q})|^2 
~2 {\cal S}(Q,\omega )
\end{equation}
where $r_0=5.38$ fm, $N$ is the number of copper ions in the sample,
$f({\bf Q})$ is the magnetic form-factor \cite{formfac}, 
and ${\cal S}(Q,\omega )$ is the
spherically averaged scattering function:
\begin{equation}
{\cal S}(Q,\omega)=\int \frac{d\Omega_{\hat{Q}}}{4\pi} \frac{1}{2}
\sum_{\alpha\beta} (\delta_{\alpha\beta}-\hat{Q}_\alpha\hat{Q}_\beta)
{\cal S}^{\alpha\beta}({\bf Q},\omega).
\label{powderavg}
\end{equation}
The dynamic spin correlation function is given by
\begin{equation}
\label{eq_sqw}
{\cal S}^{\alpha\beta}({\bf Q},\omega)=\frac{1}{2\pi\hbar}
\int dte^{i\omega t}\frac{1}{N} \sum_{{\bf R}{\bf R}^{'}}
<S^\alpha_{{\bf R}}(t)S^\beta_{{\bf R}^{'}}(0)> 
e^{-i{\bf Q} \cdot ({\bf R}-{\bf R}^{'})} .
\end{equation}

The following corrections
were applied to the neutron counts in the  time--histograms associated with
each of the three analyzer systems. First  
a time independent background measured for
$\hbar\omega\approx -E_f$ was subtracted. Then 
the data was scaled to 
the shifted time--dependent count rate in a pre--sample fission monitor, and
finally the data was converted into $\hbar\omega$--histograms.
This procedure yields $\hbar\omega$--dependent data, $I (Q,\hbar\omega )$, 
which is related to the scattering cross section
through convolution with a  resolution function as follows
\begin{equation}
I(Q,\hbar\omega )= {\cal C}
\int dQ^{\prime} \hbar d\omega ^{\prime}
{\cal R}_{Q\omega }(Q-Q^{\prime},\omega-\omega^{\prime})
\frac{k}{k^{\prime}}\frac{{\rm d}^2\sigma}{{\rm d}\Omega{\rm d}E^{\prime}}
(Q^{\prime},\hbar\omega^{\prime}),
\end{equation}
In this expression the  resolution function ${\cal R}_{Q\omega}$
is assumed to be 
normalized to unity:
\begin{equation}
\int dQ^{\prime} \hbar d\omega^{\prime} {\cal R}_{Q\omega}(
Q^{\prime},\omega^{\prime})=1
\end{equation}

For a hydrogenous sample such as CuHpCl the elastic intensity is 
dominated by the incoherent nuclear scattering cross section which reads
\begin{equation}
\frac{{\rm d}^2\sigma}{{\rm r}\Omega{\rm d}E^{\prime}}= 
N\sum_{\bf r} (b_{\bf r}^i)^2 \delta (\hbar\omega ).
\end{equation}
The summation here is over the nuclei associated with a
single copper atom.
We used the integrated elastic intensity to determine the spectrometer
constant, $N{\cal C}$: 
\begin{equation}
N{\cal C} = \frac{
\int_{-\epsilon}^{\epsilon} \hbar d\omega I(Q,\hbar\omega )}{\sum_{\bf r} 
(b_{\bf r}^i)^2}
\end{equation}
With this number we derived a normalized intensity in units suitable for
comparison to theories of ${\cal S}(Q,\omega )$,
\begin{equation} 
\tilde{I}(Q,\hbar\omega ) = \frac{I(Q,\hbar\omega )}{r_0^2N{\cal C}}
\label{Itilde}
\end{equation}
The magnetic contribution to this quantity 
is related to the spherically averaged dynamic spin correlation function 
of Eq.~(\ref{powderavg}) as follows:
\begin{equation}
\tilde{I}_m(Q,\hbar\omega ) = 2 \int dQ^{\prime} \hbar d\omega ^{\prime}
{\cal R}_{Q\omega }(Q-Q^{\prime},\omega-\omega^{\prime})
|\frac{g}{2}f({\bf Q^{\prime}})|^2{\cal S}(Q^{\prime},\omega^{\prime} )
\label{eq:ItildeQw}
\end{equation}

\label{exptech}

\section{Experimental Results}

\subsection{Magnetic Susceptibility}

Fig.\ \ref{ChiT} shows the zero-field ac susceptibility  of CuHpCl single
crystals. The characteristic rounded peak of a 1D AFM appears
at $T_p = $ 8 K.  Fitting a Curie-Weiss law to  high temperature
data (30 K $ < T < $ 200 K) gives a Curie-Weiss temperature $\Theta_{CW} =
-6.05(1)$ K.
%and g--factors of ???, and ???, along the $a$ and $b$
%directions.  
Below the peak $\chi$ drops much more rapidly than for
the linear S=1/2 chain with only NN interactions \cite{BonnerFisher},
and tends to zero at low temperatures. 
%The inset shows low temperature
%data for a powder sample. In this semi--log plot versus $1/T$ the data
%approaches a straight line for $1/T\rightarrow \infty$ indicating
%exponentially activated behavior and a spin--gap.  
These data are
consistent with previous powder susceptibility measurements
\cite{Hammar,Chaboussant}. The solid lines are a fit  based on exact
diagonalization of a spin--ladder model which  will be discussed in
detail below.

\subsection{Specific Heat}

Figure \ref{fig:CpVsT} shows the specific heat $C(H,T)$ of CuHpCl vs
$T$ in fixed fields up to $H = 9$ T.  No subtraction of the lattice 
specific heat has been attempted, but as the  data
are strongly field-dependent and  show
no sign of a contribution to $C \propto T^3$ out to $T=8$ K, the
features we observe  are evidently
dominated by the spin degrees of freedom.    
In zero field, $C$ shows a broad maximum at $T \approx 4.5 $ K,
below which it drops rapidly to zero.  As the field is increased
($H = 3$ and 5 T) the  maximum in $C$ broadens and shifts to lower
$T$, indicating a transfer of  spectral weight  to lower
energies. The qualitative low-$T$ behavior remains the same as for $H=0$,
however, as may be seen from the semi-logarithmic
plot of $C$ 
vs $1/T$ of  the $H = 0$, 3 and 5 T data shown in Fig.~\ref{fig:CpVsTinv}.
All three data sets approach straight lines in the
limit $1/T\rightarrow\infty$  indicating that $C(T)$ is
exponentially activated.  The change in slope with field 
indicates that the activation energy 
decreases with increasing $H$.  This may be quantitatively related to the
closure of the spin gap with field, as will be discussed in 
Sect.~\ref{sec:CpAnalysis} below.  

As $H$ is increased further, the shift of spectral weight to lower
energy  continues as  shown in Fig.~\ref{fig:CpVsT}.  As the critical
field $H_{c1} \simeq 7$ T \cite{Hammar,Chaboussant} 
where the gap closes is passed, the curvature of the low--$T$ data
changes, and first a shoulder and then a second broad peak appear
at $T \approx 1$ K.  Above $H_{c1}$, a cusp-like anomaly also becomes 
visible on top
of this second peak.  These low--$T$ features are shown  in more detail in
Fig.~\ref{fig:CpVsTsqr}, where we plot $C /T$ vs $T^2$ for $T < 1.4$ K.
The cusp can clearly be seen at $H = 8$ T and $H = 9$ T, and below the
cusp,  $C(T) \propto T^{3}$. These features
suggest that above $H_{c1}$ a phase transition occurs to a
field-induced  ordered state. A cusp is not as clearly visible
at $H = 7.5$ T, but below the pronounced knee in that data set
$C(T) \propto T^{3}$, indicating that at this field the system
is also ordered.  In contrast, for $H = 6.6$ T, at the lowest temperatures
shown $C$ drops faster than $T^3$, suggesting the existence of
a small gap \cite{GapAt66T}. The inset to 
Fig.~\ref{fig:CpVsTsqr} shows the $H$--$T$ phase
diagram of CuHpCl as determined from these data in combination with our earlier
field-dependent susceptibility measurements \cite{Hammar}, and the
magnetization data of Chaboussant {\em et al.} \cite{Chaboussant}.
We note that this phase diagram, and indeed the overall features of the
susceptibility and field-dependent specific heat are quite similar to
other gapped spin chains, such as the alternating chain
$\rm Cu(NO_3)_2\cdot 2.5 H_2O$ \cite{Bonner83}.

\subsection{Neutron Scattering}

Figure \ref{Sw} shows the normalized angular average 
$\tilde{I}(\hbar\omega )$ of neutron
scattering from CuHpCl. 
The normalization of $\tilde{I}(\hbar\omega )$ is as 
described in Eq.~\ref{Itilde}, and
the angular average was carried out over the full accessible range
of scattering angle $21^o < 2 \theta < 153^o$.
The intense elastic peak arises from incoherent
nuclear scattering as discussed in Sect.~\ref{exptech} and the
inelastic scattering for $\hbar\omega > 2$ meV comes from incoherent
inelastic nuclear scattering. There is however also a broad peak for 0.8 meV
$<\hbar\omega <$ 1.4 meV which we now show is
associated with magnetic scattering.
To isolate this feature we measured the incoherent nuclear scattering
from 5.4 g of the non-magnetic organic compound benzophenone (dibenzyl
ketone).  The material and the amount were chosen
 to match the total elastic cross
section of  CuHpCl, and filled approximately the same 
volume as the CuHpCl sample.  
33 mg of boron carbide was added to the benzophenone to
match the absorption cross-section of CuHpCl.  The resulting integrated
elastic intensity from the benzophenone sample was only 15\% stronger
than for CuHpCl. We applied a scale factor to the
benzophenone  data so that its integrated elastic
intensity matched that of $\tilde{I}(Q,\hbar\omega )$ for CuHpCl. The
angular-averaged
data for benzophenone are shown as open squares in Fig.~\ref{Sw}. As may
be seen, the line shapes
of the elastic peaks are indistinguishable, which allows us to
use the benzophenone 
data to approximately subtract the non-magnetic contributions to 
the neutron scattering intensity from CuHpCl. A small constant
term of $0.03$ meV per Cu was added to ensure that the difference data
for $\hbar\omega << -k_B T$ averages to zero as required by detailed balance.

The resulting difference data, $\tilde{I}_m(\omega)$, is shown in 
Fig.~\ref{dIw}(a).  The only feature in these data above background is 
the peak centered at $\hbar\omega = 1.15$ meV. To determine whether 
this feature is magnetic or is due to phonons, we measured the 
scattering from the CuHpCl sample at $T=20$ K and subjected those data 
to the same background corrections as the low temperature data.  The 
result is shown in Fig.~\ref{dIw} (b).  The peak broadens and a mirror 
image of it emerges for $\hbar\omega <0$ as expected from detailed 
balance\cite{lovesey}.  This behavior establishes that the difference 
data shown in Fig.~\ref{dIw}(a) is indeed magnetic and is a measure of 
the the angular averaged magnetic scattering intensity for CuHpCl:
\begin{equation}
\tilde{I}_m(\hbar\omega ) = 
\int_{21^o}^{153^o} d2\theta \tilde{I}_m(Q,\hbar\omega ) ,
\end{equation}
which may related to  ${\cal S}(Q,\omega )$ through
Eq.~\ref{eq:ItildeQw}.
The data shows that magnetic scattering occurs only above a gap,
$\Delta\approx 0.8$ meV and that the bandwidth for the magnetic excitation
spectrum is 0.7 meV. 

Having established the magnetic ``density of states,'' we turn to the 
wave vector dependence of the magnetic scattering cross section which 
contains information about spatial spin correlations. In Fig.~\ref{Qdep}
we show the $Q$--dependence of the magnetic scattering integrated over the
range of energies with appreciable magnetic signal: 
\begin{equation}
\tilde{I}_m(Q) = 
\int_{0.8~meV}^{1.5~meV} \hbar d \omega \tilde{I}_m(Q,\hbar\omega ) ,
\label{eq:ItildeQ}
\end{equation}
 The limited sampling density and statistics
do not allow us to clearly discern any features in the data. We note 
only that as expected for magnetic scattering and not for inelastic 
scattering from phonons the intensity decreases with increasing wave vector 
transfer. The lines through the data in Figs.~\ref{dIw} and \ref{Qdep}
result from a model calculation to
be discussed below.

\section{Discussion}

\subsection{Analysis of susceptibility data}
In their pioneering work on CuHpCl \cite{Chiari90}, Chiari {\em et al.} 
suggested that this material may realize the Hamiltonian ${\cal H}$ 
of Eq.~(\ref{H_NNN}).  On the basis of high temperature susceptibility 
data alone, they were not able to place strong constraints on the 
possible models, although they concluded 
that the intra-dimer and inter-dimer NN interactions $J_1$ and $J_3$ 
are different as indicated in Fig.~\ref{structure}. 

We have previously analyzed some of the data on this material in terms
of the isolated dimer model ( Eq.~(\ref{H_NNN}) with $J_{3}=0,J_2=0$),
the alternating near-neighbor model ($J_2 = 0$, $J_{3}/J_{1} \in
[0,1] $), and the next-nearest-neighbor model without alternation ($
J_{3}/J_{1} = 1$, $J_2/J_{1} \in [0.3,0.7] $).  None of these
provided a satisfactory fit to our powder susceptibility data
\cite{Hammar}.  Chaboussant {\em et al.} analyzed their powder
susceptibility data in terms of both the alternating chain and the
spin ladder ($J_3 = 0$) models, using high temperature series
expansions \cite{Chaboussant}.  While both models fit the data, from
structural considerations they concluded that the ladder is more
likely.  A comparison of their magnetization data to exact
diagonalization calculations for the ladder model \cite{Hayward96}
found good agreement for $J_2/J_1 = 0.18$, although additional
improvement to the fit was found if a small ferromagnetic diagonal
coupling $J_3$ was added.

To analyze our single crystal susceptibility data, we 
carried out a  
systematic search of the parameter space
of the Hamiltonian in Eq.~(\ref{H_NNN}). We  
diagonalized chains up to length N = 14 for values of the exchange 
constants  
on a large grid in $J_2/J_1-J_3/J_1$ space.  The calculated eigenvalues for 
each point were used to generate $\chi(T/J_1,H=0)$ on the interval 
$T/J_1 \in [0.2,20]$.  Below $T/J_1 = 0.2$, the results for N = 12 
deviated from those for N = 14 by greater than 1\%, due
to effects of the finite system size.  By tri--linearly 
interpolating in this 3D mesh of $(T/J_1,J_2,J_{3})$, we could
generate $\chi(T)$ for 
arbitrary $J_1,J_2,$ and $J_{3}$.  The accuracy of this procedure was 
checked by comparing the interpolated $\chi(T)$ between mesh points 
against exact diagonalization at those same values.  The difference 
was found to be less than 1 part per thousand.  The parameter space 
was constrained by the measured upper critical field
$H_{c_2} = 13.3$T, or 18.8K
\cite{Chaboussant}, which occurs at 
$g\mu_BH_{c2} =\max{(J_1+J_{3},J_{1}+2*J_2))}$ where the energy of the 
spinwaves from the ferromagnetic state change sign.  

A simultaneous fit to both the a-axis and b-axis data in
Fig.~\ref{ChiT} gave
%, using the measured g-factors \cite{Chaboussant}, gave 
$g_a = 2.15(1)$, $g_b = 2.04(1)$, 
$J_1/k_B = 13.26(4)$ K, $J_2/k_B = 2.4(3)$ K, and $J_3/k_B = 1.1(6)$ K.
The error bars quoted correspond to the intervals over which 
the reduced $\chi^2$ deviates by less than $1/N_{free}$ from its minimum
value. Here
$N_{free}$ is the number of data points minus the number of parameters
in the fit.
The results are shown as solid curves in the figure.
The values of the g-factors in our fits are consistent with the values
determined by EPR,\cite{Chaboussant} and the
values for $J_1$ and $J_2$  are in good agreement with the
analysis of the magnetization data, although $J_3$ on which 
our error bars are large differs in sign.

The fit indicates that  the intra rung coupling  $J_1$ exceeds the
inter--rung coupling $J_2$  by a factor of approximately five.
In this strong--coupling 
limit the zeroth order approximation to the ground state is simply 
a chain of non--interacting rung singlets and the excited states are
localized triplets at energy $\hbar\omega = J_1$. Clearly this approximation is
too coarse since it cannot account for the finite bandwidth of the 
magnetic excitation spectrum which is evident in 
the neutron scattering data (Fig.~\ref{dIw} (a)). 
We therefore must take the inter-rung coupling into account, and the 
most straightforward way to do this is 
through a
perturbation expansion. 
Neglecting effects of $J_3$, 
the excited states then become propagating triplets with
a dispersion relation\cite{reigrotzki}
\begin{eqnarray}
\hbar\omega (\tilde{q}) &=& J_1 [1+\frac{J_2}{J_1}\cos \tilde{q}
+\frac{1}{4}(\frac{J_2}{J_1})^2(3-\cos 2\tilde{q})\nonumber\\
&&-\frac{1}{8}(\frac{J_2}{J_1})^3(2\cos \tilde{q}+2\cos 2\tilde{q}
-\cos 3\tilde{q} -3)+O(\frac{J_1}{J_2})^4] ,
\label{disprel}
\end{eqnarray}
where $\tilde{q} = {\bf Q} \cdot {\bf u}$ is the component of 
wavevector transfer along the chain.
For our ladder model the higher order terms in  
Eq.~({\ref{disprel}) are very small. 
For example $J_1(J_2/J_1)^2=0.039$ meV which could be 
compared to the energy resolution of our neutron scattering 
experiment  which is $\delta E=0.12$ meV FWHM.
Therefore we approximate the dispersion relation with 
\begin{equation}
\hbar\omega ({\bf Q}) = J_1 + J_2 \cos ({\bf Q}\cdot {\bf u}) .
\label{wdisp}
\end{equation}
The corresponding gap in the magnetic excitation spectrum is 
\begin{equation}
\Delta = J_1-J_2
\end{equation}

\subsection{Analysis of Specific Heat Data}
\label{sec:CpAnalysis}
To test  the spin ladder model we
compare its predictions
 to the low temperature specific heat data. 
If we approximate the dispersion relation 
by Eq.~(\ref{wdisp}) and neglect interactions between propagating triplet
excitations, the zero-field
specific heat in the low temperature limit takes the form
\begin{equation}
C = \frac{n R}{2 \sqrt{2\pi}} \left (\frac{ \Delta}{k_B T}\right)^{3/2}
     \sqrt{\frac{\Delta}{J_2}}\exp(-\Delta/k_B T) ,
\label{CpEqn}
\end{equation}
measured in J/(K mol-Cu) with
$n=3$ being the number of low energy modes in a Brillouin zone. 
The solid line through the zero-field data
in Fig.~\ref{fig:CpVsTinv} is a fit to  this functional 
form. It yields
$\Delta =   0.89(1)$ meV, which is close to the value 
$\Delta = J_1-J_2=0.93 (3)$ meV  predicted by the spin ladder model.
If we assume that the main effect of fields $H < H_{c1}$ on the spectrum is
to Zeeman-split the triplet mode, then Eq.~\ref{CpEqn} should remain a
good description of the low-$T$ behavior of $C(H,T)$, although
both $n$ and $\Delta$ may now vary with field.  The fits shown in 
Fig.~\ref{fig:CpVsTinv} for $H = 3 $ T and 5 T yield $\Delta = 0.55(4)$ meV
and $n = 1.3(3)$, and $\Delta = 0.24(4)$ and $n = 0.6(2)$, respectively.
This reduction of $n$ with field is qualitatively
consistent with  spin-splitting of the triplet, as the 
lowest-energy band should dominate the specific heat in the 
range $k_B T < \Delta (H)/4$ where these fits were done.
The   gap vs field determined from the fits is
shown in the inset of Fig.~\ref{fig:CpVsTinv} together with the
estimated value of the gap at $H = 6.6$ T \cite{GapAt66T}.  
The observed reduction of $\Delta$ with $H$ is consistent with
the linear behavior expected for the Zeeman-splitting of the triplet.
The data in the inset are fit by the line shown, which has a
slope of $-2.09(1) \mu_B$, and extrapolates to
$\Delta = 0$ at $H = 7.2(1) $ T, 
in accord with $H_{c1} $ determined by other 
methods\cite{Hammar,Chaboussant}.

\subsection{Single mode approximation to ${\cal S}(Q,\omega )$}
A spin ladder Hamiltonian with  $J_3 \approx 0$ thus can account for
the temperature dependent susceptibility and the low temperature specific heat
of CuHpCl.  
We now discuss to what extent the spin ladder model can account for 
our neutron scattering data. 

Exact diagonalization studies suggest that most of the spectral weight
for magnetic excitations is associated with the resonant 
triplet mode \cite{Barnes93}.
Therefore we use the single mode approximation: that the dynamic spin 
correlation function takes the form
\begin{equation}
	{\cal S}^{\alpha\beta}({\bf Q},\hbar\omega )=
	{\cal S}({\bf Q})\delta (\hbar\omega -\hbar\omega ({\bf 
	Q})) \delta_{\alpha,\beta} ,
	\label{SMAa}
\end{equation}
Two sum--rules\cite{Hohenberg}
 place further constraints on ${\cal S}({\bf Q},\omega )$:
\begin{eqnarray}
\int d^3{\bf Q} \int \hbar d\omega \sum_{\alpha}{\cal S}^{\alpha,\alpha}({\bf Q},\omega )/\int d^3{\bf Q}
&=&S(S+1)
\label{totalmoment}\\
\hbar^2\int \omega d\omega {\cal S}^{\alpha\beta}({\bf Q},\omega )
&=&<[[{\cal H},S^{\alpha}_{\bf Q}],S^{\beta}_{-\bf Q}]>,
\label{firstmoment}
\end{eqnarray}
Where 
\begin{equation}
S^{\alpha}_{\bf Q} = \frac{1}{N}\sum_{\bf R} S^{\alpha}_{\bf R} e^{-i{\bf Q}
\cdot {\bf R}}.
\end{equation}
From Eqs. (\ref{firstmoment}), (\ref{SMAa}), and (\ref{H_NNN})  
with $J_3 = 0$
it can be shown that
\begin{equation}
	{\cal S}({\bf Q})= -{2 \over 3}{1 \over { \hbar\omega ({\bf Q})}}\left[ 
	\matrix{J_1\left\langle {{\bf S}_{2i}\cdot {\bf S}_{2i+1}} 
		\right\rangle \left( {1-\cos ({\bf Q}\cdot {\bf d})} \right)\hfill\cr
		  +J_2 \left( \left\langle {{\bf S}_{2i}\cdot {\bf S}_{2i+2}} \right\rangle 
	  +\left\langle {{\bf S}_{2i-1}\cdot {\bf S}_{2i+1}} \right\rangle \right)
	  \left( {1-\cos ({\bf Q}\cdot {\bf u})} \right)\hfill\cr} \right]
	\label{SMAb}
\end{equation}
Here ${\bf d}$ and ${\bf u}$ are vectors connecting spins within a rung
and along an arm of the ladder, respectively, and 
$\left\langle {{\bf S}_{i}\cdot {\bf S}_{j}} \right\rangle$ are
the corresponding groundstate equal time correlation functions.
Because we are interested in the strong rung coupling limit we
neglect interdimer correlations and
adopt a simplified version of Eq.~(\ref{SMAb})
\begin{equation}
	{\cal S}({\bf Q}) \approx -{2 \over 3}\frac{J_1}  
	{ \hbar\omega ({\bf Q} \cdot {\bf u})} 
	\left\langle {{\bf S}_{2i}\cdot {\bf S}_{2i+1}} \right\rangle 
	\left( {1-\cos ({\bf Q}\cdot {\bf d})} \right) .
	\label{SMAc}
\end{equation}
In this approximation the coupling along the ladder is reflected only
in the one-dimensional dispersion relation 
$\hbar\omega ({\bf Q} \cdot {\bf u})$ of Eq.~(\ref{wdisp}).
Requiring that ${\cal S}({\bf Q},\omega)$ defined in this way satisfies
the total moment sum rule of Eq.~(\ref{totalmoment}) fixes
\begin{equation} 
\left\langle {{\bf S}_{2i}\cdot {\bf S}_{2i+1}} \right\rangle 
 = -S(S+1)(1 - (J_2/J_1)^2)^{1/2}.
\end{equation}
With the exchange
constants determined from the susceptibility measurements,
this  gives a model that may
be compared to the data of Figs.~\ref{dIw} and \ref{Qdep} 
with no adjustable parameters.

After inserting Eq.~(\ref{SMAc}) in Eq.~(\ref{powderavg}),
the spherical average may be computed exactly, yielding
\begin{eqnarray}
{\cal S}(Q,\hbar\omega) &=& 
-\frac{2 J_1 \left\langle {{\bf S}_{2i}\cdot {\bf  S}_{2i+1}} 
\right\rangle}{3 Q u} \sum_{Q_i \le Q}
\frac{1- J_0(d \sin \alpha \sqrt{Q^2 - Q_i^2}) \cos (Q_i d \cos \alpha)}
      {\hbar\omega \sqrt{(J_1 + J_2 -\hbar\omega)(\hbar\omega - (J_1 - 
      J_2))}} \nonumber \\
& &   \times \Theta (J_1 + J_2 -\hbar\omega) \Theta (\hbar\omega -(J_1 - 
J_2))
\label{powderavg2}
\end{eqnarray}
where $J_0 (x)$ is a Bessel function, $\alpha = 55.4^\circ$ is the angle
between ${\bf u}$ and ${\bf d}$,\cite{Chiari90} and the 
$Q_i = (1/u) \cos^{-1}[(\hbar\omega - J_1)/J_2]$
are the solutions of  Eq.~(\ref{wdisp}) at fixed $\hbar\omega$.
Convolution of Eq.~(\ref{powderavg2}) with the instrumental resolution
and summing over the values of $2\theta$ used in the experiment
gives the solid curve in 
Fig.~\ref{dIw}(a).  Several features of this curve are worth noting.
First, the mean energy and bandwidth match the data quite well.
This is to be expected since the model after all was chosen to 
account for susceptibility data whose temperature dependence 
controlled by these two features of the magnetic excitation spectrum.
Second the area under the data closely matches the 
area under the calculated curve. This indicates that the inelastic
peak which we have detected contains most of the magnetic neutron scattering
to be expected from the copper ions in CuHpCl.

The model does not, however, reproduce the shape of the data very well.
The two peaks in the model correspond to singularities in the 
one-dimensional magnetic density of states at $\hbar\omega = J_1 \pm J_2$.
The data does not show these peaks, being instead peaked near the center
of the band.  Note that inclusion of the interdimer correlations from 
Eq.~\ref{SMAb}, which were neglected in our approximate treatment 
(Eq.~\ref{SMAc}),
will only make the agreement of model with data worse, as that produces
a contribution to $S(Q,\omega)$ with a single peak at $\hbar\omega = J_1 
- J_2$. We also note that the data show an unexpected tail of scattering
intensity below the lower bound predicted by the model.
This is consistent with  specific heat data \cite{Hammar}
that show a gap $\Delta = 0.89(1)$ meV,
which is close to the gap seen from neutron scattering data
but less than the gap $\Delta = J_1 - J_2 = 0.93(3)$ meV
predicted by the  ladder model derived from the susceptibility data.

Fig.~\ref{Qdep} compares the data and the model calculation of the
$Q$--dependent energy integrated intensity (Eq.~\ref{eq:ItildeQ}).
For CuHpCl, this is the first
comparison of data which is sensitive to the geometry of the spin
system to predictions of the ladder model. 
The data bear no evidence of the maximum in $\tilde{I}(Q)$ predicted by the
spin ladder model for  $Q\approx 1.3$\AA$^{-1}$, which 
raises the   question of whether  the ladder model is the correct description 
of CuHpCl.

\section{Conclusion}

We have reported experiments on CuHpCl which establish that this system
has a gap in its magnetic excitation spectrum and which determine the 
magnetic bandwidth. Our susceptibility and specific heat data can 
be accounted for in terms of a spin--ladder model with 
intra--rung coupling $J_1 = 1.143(3)$ meV and two mutually frustrating
inter--rung interactions:  $J_2 = 0.21(3)$ meV and $J_3 = 0.09(5)$
meV. Not surprisingly this model reproduces the gross features of the
magnetic density of states as probed by neutron scattering. However 
our neutron scattering data do not show the Van--Hove singularities 
and the Q--dependence predicted by a single mode approximation 
of the excitation
spectrum 
in a simple spin ladder model.  It is possible that
a refined calculation of ${\cal S}(Q,\omega )$ for a spin ladder would resolve 
these discrepancies, but it is clear that further experiments, particularly
neutron scattering on 
deuterated single crystals, are required to definitively establish the 
spin Hamiltonian for CuHpCl.

\acknowledgements
This work was supported by the NSF under grants DMR-9357518 and 
DMR-9302065, the DOE BES-Materials Science contract W-31-109-ENG-38 
with IPNS-ANL, and by the David and Lucile Packard Foundation.

\begin{figure}
\caption{ Schematic of chain structure of CuHpCl, indicating possible 
near-neighbor bond alternation ($\delta$) and next-near-neighbor bonds. 
The ellipses show how the dimer units stack. } 
\label{structure}
\end{figure}

\begin{figure}
\caption{
AC susceptibility $\chi (T)$ of single crystal CuHpCl with the field
oriented along the ${\bf a}$ and ${\bf b}$ axes.  The 
curves differ by a scale factor consistent with the g-factor 
anisotropy.\protect\cite{Chaboussant} 
The solid lines are a fit to exact 
diagonalizations of the $J_1$-$J_2$-$J_3$ model.
}
\label{ChiT}
\end{figure}

\begin{figure}
\caption{
Magnetic specific heat $C(H,T)$ vs  $T$ for CuHpCl.  
At low fields $C$ is suppressed at low temperatures because of the 
gap to magnetic excitations.  Above $H_{c1} \approx 7$ T, there is
evidence for a phase 
transition to an ordered state.
}
\label{fig:CpVsT}
\end{figure}

\begin{figure}
\caption{
Semi-logarithmic plot of $C(H,T)$ vs $1/T$ in the gapped phase
showing activated behavior with fits to a spin-ladder model
as described in the text.
Inset: gap $\Delta$ determined from fits plus an estimate of gap
for $H = 6.6$ T \protect\cite{GapAt66T}.
}
\label{fig:CpVsTinv}
\end{figure}

\begin{figure}
\caption{
$C/T$ vs $T^2$ at low temperatures, showing $C \propto T^3$ 
below the temperature of the ordering anomaly at high fields. 
Inset: putative phase diagram for CuHpCl based on specific heat
and magnetization data\protect\cite{Hammar,Chaboussant}, indicating
boundary between one-dimensional (1D) disordered and ordered (AF) phases.
}
\label{fig:CpVsTsqr}
\end{figure}

\begin{figure}
\caption{
Normalized, angular-averaged inelastic neutron scattering 
intensity for CuHpCl at $T = 1.5$ K (solid circles).  The full incoherent
elastic peak (small circles) is shown reduced in magnitude 1000 times.
Also shown is scattering intensity for benzophenone (open squares), used
to measure the non-magnetic background for -1.5 meV $\le \hbar\omega \le$
2 meV.
}
\label{Sw}
\end{figure}

\begin{figure}
\caption{ 
Magnetic, angular-averaged inelastic neutron scattering 
intensity for CuHpCl obtained after subtracting non-magnetic
background in Fig. 6.  The reduction of the peak in (a) with increasing $T$
identifies it as due to spin fluctuations.  The solid line in (a) is
the predicted intensity based on the spin ladder model derived from 
susceptibility measurements with no adjustable parameters.
}
\label{dIw}
\end{figure}

\begin{figure}
\caption{
 Wavevector dependence of the energy-integrated magnetic 
scattering for CuHpCl at $T = 1.5$ K.  Solid line is the prediction 
of the spin ladder model.
}
\label{Qdep}
\end{figure}


\begin{references}
\bibitem{Bonner79} J. C. Bonner, H. W. Bl\"{o}te, J. W. Bray,
I. S. Jacobs, J. Appl. Phys. {\bf 50}, 1810 (1979)

\bibitem{Bonner83}
    J. C. Bonner, S. A. Friedberg, H. Kobayashi, D. L. Meier, and
    H. W. J. Bl\"ote, Phys. Rev. B {\bf 27}, 248 (1983).
\bibitem{Shastry81}
B. S. Shastry and B. Sutherland, Phys. Rev. Lett. {\bf 47}, 964 (1981).

\bibitem{MajumdarGhosh69}
C. K. Majumdar and D. K. Ghosh, J. Phys. C {\bf 3}, 911 (1969).

\bibitem{white}S. R. White and I. Affleck, Phys. Rev. B {\bf 54}, 9862 (1996).
\bibitem{dagotto}E. Dagotto and T. M. Rice, Science {\bf 271}, 618 (1996).
\bibitem{LadderTheory}
    S.  Gopalan, T.  M.  Rice, and M.  Sigrist, Phys.  Rev.  B {\bf 49}, 
    8901 (1994); S.  R.  White, R.  M.  Noack, and D.  J.  Scalapino, 
    Phys.  Rev.  Lett {\bf 73}, 886 (1994); S.  P.  Strong and A.  J.  
    Millis, Phys.  Rev.  Lett {\bf 69}, 2419 (1992); M.  Azzouz, L.  Chen, 
    and S.  Moukouri, Phys.  Rev.  B {\bf 50}, 6233 (1994).
\bibitem{Eckert}J. Eckert, D. E. Cox, G. Shirane, S. A. Friedberg, and 
H. Kobayashi, Phys. Rev. B {\bf 20}, 4596 (1979).

\bibitem{arai}M. Arai, M. Fujita, M. Motokawa, J. Akimitsu, and S. M. Bennington, Phys. Rev. Lett. {\bf 77}, 3649 (1996).
\bibitem{ain}M. Ain,J. E.  Lorenzo, L. P.  Regnault, G.  Dhalenne,
A. Revcolevschi, B.  Hennion, and T. Jolicoeur, Phys. Rev. Lett.
{\bf 78}, 1560 (1997).
\bibitem {Azuma94}
    M. Azuma, Z. Hiroi, M. Takano, K. Ishida, and Y. Kitaoka,
    Phys. Rev. Lett. {\bf 73}, 3643 (1994).

\bibitem{Carter96}
S. A. Carter, B. Batlogg, R. J. Cava, J.J. Krajewski, W. F. Peck, Jr., and
T. M. Rice, Phys. Rev. Lett. {\bf 77}, 1378 (1996).
\bibitem{Chiari90} B.\ Chiari, O.\ Piovesana, T.\ Tarantelli, and P.\
F.\ Zanazzi,  Inorg. Chem. {\bf 29}, 1172 (1990).

\bibitem{Hammar}
P. R. Hammar and  D. H. Reich, J. App. Phys. {\bf 79}, 5392 (1996)

\bibitem{Chaboussant}
G. Chaboussant, P. Crowell, L. P. L\'{e}vy, O. Poivesana, A. Madouri, D. 
Mailly, Phys. Rev. B {\bf 55}, 3046 (1997).
\bibitem{cuhpclnmr}
G. Chaboussant, M.-H. Julien, Y. Fagot-Revurat, L. P. Levy, C. Berthier,
M. Horvatic, and O. Piovesana, Preprint cond-mat/9706138 (1997).

\bibitem {Johnston87}
    D. C. Johnston, J. W. Johnson, D. P. Goshorn, and  A. J. Jacobson,
    Phys. Rev. B {\bf 35}, 219 (1987).
\bibitem{eccleston}R. S. Eccleston, T. Barnes, J. Brody, and J. W. Johnson,
Phys. Rev. Lett. {\bf 73}, 2626 (1994).
\bibitem{garrettprb}A. W. Garrett, S. E. Nagler, T. Barnes, and B. C. Sales,
Phys. Rev. B {\bf 55}, 3631 (1997).  
\bibitem{garrettprl}A. W. Garrett, S. E. Nagler, D. A. Tennant, B. C. Sales,
and T. Barnes, Phys. Rev. Lett. {\bf 79}, 745 (1997).
\bibitem{chitra97}
R. Chitra, and T. Giamarchi, Phys. Rev. B {\bf 55},  5816 (1997).

\bibitem{Bachmann72} R.\ Bachmann, {\em et al.},
Rev.\ Sci.\ Inst.\ {\bf 42}, 205 (1972).

\bibitem{lovesey} We use the notation of
S.W. Lovesey, {\it Theory of Neutron Scattering
from Condensed Matter}, Clarendon Press., Oxford (1984).

\bibitem{formfac}A. J. Freeman and R. E. Watson, Acta. Cryst. 
{\bf 14}, 231 (1961); P. J. Brown, in {\em International Tables 
for Crystallography}, A. J. C. Wilson Ed. , Vol. C, Kluwer Academic Publishers,
London (1995).
\bibitem{BonnerFisher}
J. C. Bonner and M. E. Fisher, Phys. Rev. {\bf 135}, A640 (1964).

\bibitem{GapAt66T}
    We estimate the gap $\Delta(H=6.6$ T) $\approx 0.10(5)$ meV based
    on experience with field-induced
    gaps in other nearly-critical spin chain systems
    (Dender {\em et al.}, Phys.  Rev. Lett, to appear) that shows the
    ``knee" in a plot of $C/T$ to reliably occur at $\approx \Delta/4$.
\bibitem{Hayward96}
C. A. Hayward, D. Poilblanc, and  L. P. L\'{e}vy, Phys. Rev. B {\bf 54},
R12649 (1996)

\bibitem{reigrotzki}
M. Reigrotzki, H. Tsunetsugu, and T. M. Rice, J. Phys. Condens. Matter 
{\bf 6}, 9235 (1994).

\bibitem{Barnes93}
    T. Barnes, E. Dagotto, J. Riera, and E. S. Swanson, Phys. Rev. B {\bf 
    47}, 3196 (1993);
\bibitem{Hohenberg}
     P. C. Hohenberg and W. F. Brinkman, Phys. Rev. B {\bf 10}, 128 (1974).
\end{references}
\end{document}